\documentclass[11pt]{article}

\usepackage[T1]{fontenc}
\usepackage[utf8]{inputenc}
\usepackage{lmodern}
\usepackage{amsmath,amssymb,amsthm,mathtools}
\usepackage{booktabs}
\usepackage{enumitem}
\usepackage{hyperref}
\usepackage{cleveref}
\usepackage[a4paper,margin=1in]{geometry}
\usepackage{authblk}
\usepackage[numbers,sort&compress]{natbib} 

\hypersetup{
  colorlinks=true,
  linkcolor=blue,
  citecolor=blue,
  urlcolor=blue
}

\usepackage{algorithm}
\usepackage{algorithmic}
\usepackage[utf8]{inputenc}
\usepackage{textcomp}
\usepackage{tikz}
\usetikzlibrary{positioning,arrows.meta}

\title{Risk‑Averse Alert Prioritization for IDS Using Subnormal Gaussian Fuzzy Models}

\author{Murat Moran}
\affil{Department of Computer Engineering, Giresun University, T\"{u}rkiye}
\affil[ ]{\texttt{murat.moran@giresun.edu.tr}} 
\date{}

\begin{document}

\maketitle


\begin{abstract}
Modern intrusion detection systems generate thousands of alerts daily, but alert fatigue severely limits security operations effectiveness due to too many false positives or low-impact events. We address this by proposing a principled framework for alert prioritization based on subnormal Gaussian fuzzy numbers, explicitly modeling three sources of uncertainty: threat severity, detection confidence, and organizational risk attitude. Each alert is represented as a fuzzy number with the core indicating severity, spread indicating uncertainty, and height reflecting detection reliability. We apply ranking indices to prioritize alerts, allowing organizations to tune security posture through a risk-attitude parameter. Experimental validation on CIC-IDS2017 and NSL-KDD demonstrates greater robustness than baselines under detector degradation (0.9963 vs 0.8215 NDCG$_{\text{rel}}$@100), with distinct differentiation in mid-confidence alerts and near-parity with baselines under robust detectors. The framework is theoretically grounded, computationally efficient, provides interpretable reasoning, and remains robust across detector families and miscalibration scenarios.
\end{abstract}

\vspace{0.3cm} 
\noindent \textbf{Keywords:} Intrusion Detection System, Alert Prioritization, Risk-Averse Decision-Making, Alert Fatigue, Fuzzy Logic, Cybersecurity Operations, Uncertainty Quantification

\section{Introduction}

Modern intrusion detection systems (IDS) are essential components of the enterprise cybersecurity infrastructure. However, their effectiveness is severely limited by a critical operational challenge: \textit{alert fatigue}, i.e., reduced analyst responsiveness caused by excessive security alerts. This issue has been extensively studied in the context of security operations centers (SOCs), where it is recognized as a significant contributor to the failure to detect attacks~\cite{Wang2024alertfatigue}. SOCs are typically busy with thousands of alerts daily; however, the overwhelming majority are false positives (FPs) or low-severity incidents that do not require immediate attention. Consequently, this high volume of alerts consumes analysts' ability to prioritize effectively, which in turn results in delayed detection of critical threats, analyst burnout, and substantial operational expenses.

The scale of this problem is substantial. Recent surveys indicate that a typical enterprise IDS generates 10,000 or more alerts every day, with more than 50\% being FPs~\cite{Jalalvand2024alert}. This volume of alerts causes significant alert fatigue among security analysts~\cite{Alahmadi2022false}, with 64\% of SOC teams reporting being overwhelmed by FPs~\cite{Sans2024detection}. Consequently, analysts often ignore alerts, disable rules, or pass them to colleagues~\cite{Jalalvand2024alert}, potentially missing critical threats. This inefficiency has direct business consequences: increased incident response times, higher breach detection latency, wasted analyst resources, and ultimately reduced security posture. In this context, the ability to accurately rank and prioritize alerts is not merely a convenience, but a necessity for effective cybersecurity.


The core problem of alert fatigue is not that IDS fails to find attacks. Instead, it comes from the large number of alerts generated by current systems, along with a lack of effective prioritization. This prioritization issue is essentially a problem of \textit{decision making under uncertainty}. Each alert has several related sources of uncertainty that current systems do not handle well. These include:

\textbf{Severity Uncertainty:} The risk associated with different attacks varies. A vulnerability within a critical system presents a greater risk than one in a non-critical system. Nevertheless, the assessment of severity is itself uncertain due to several factors: 

\begin{enumerate}
    \item The actual impact depends on the organization's specific infrastructure and business context
    \item Standard severity metrics, such as the Common Vulnerability Scoring System (CVSS), are generic and may not reflect organizational risk
    \item Novel attacks may not have established severity assessments
\end{enumerate}

\textbf{Detection Confidence Uncertainty:} Not all alerts are equally reliable. The confidence that an alert represents a true attack depends on the detection method (feature set, training data, and model choice), the attack class, the specific network context, and whether the attack is novel/zero-day (where validation evidence is limited).

\textbf{Classification Uncertainty:} When an IDS flags an alert, it is not always clear what kind of attack is actually taking place: it could be a SQL injection, a buffer overflow, or something entirely different. This uncertainty then affects how we assess the severity of the threat, which in turn impacts how we prioritize our responses.

\subsection{Limitations of Existing Approaches}
Current alert ranking methods treat these uncertainties implicitly or ignore them entirely. They produce single-point estimates (e.g., a severity score or confidence score) without quantifying the uncertainty around those estimates. This leads to suboptimal prioritization: high-severity but low-confidence alerts (potential zero-days) may be ranked below low-severity, high-confidence alerts, or vise-versa, depending on the ad-hoc weighting scheme used. A principled approach to alert prioritization must explicitly model and reason about these multiple sources of uncertainty.

There are different ways to prioritize alerts, but each has its own set of problems. \textit{Severity-based ranking} uses common metrics such as CVSS scores. But it does not take into account how sure the detection is, which means that analysts might waste time on alerts that aren't very reliable, even if they are very serious.

Conversely, \textit{Confidence-based ranking} emphasizes alerts generated by reliable detection techniques. However, this method overlooks the severity, which could result in the omission of crucial attacks that are more challenging to identify with precision. An alternative strategy involves \textit{Ad-hoc weighted combinations}. These methods attempt to reconcile severity and confidence through the application of a straightforward weighted sum or product. However, the weights are arbitrarily chosen, without any theoretical basis, and do not reflect the organization's specific risk preferences. Lastly, \textit{Machine learning approaches} learn prioritization from historical data, but are typically black-box, lack interpretability, and require expensive labeled training data~\cite{Sommer2010}.

As a result, none of these approaches explicitly model the uncertainty inherent in severity and confidence assessments. They treat these quantities as point estimates rather than uncertain quantities with distributions and confidence intervals. As a result, they cannot distinguish between high-confidence estimates and low-confidence estimates, nor can they adapt their decision-making based on organizational risk tolerance.

\subsection{Fuzzy Logic and Risk-Averse Decision-Making as a Solution}

We present a natural, theoretically sound framework for managing multiple sources of uncertainty in alert prioritization, employing fuzzy logic. Unlike probabilistic methods, which require exact probability distributions, fuzzy logic is capable of processing incomplete data and expert evaluations. \textit{Subnormal fuzzy sets}, which have a maximum membership degree less than 1, are particularly useful for this problem. They naturally represent the idea of partial reliability. For example, a high-severity alert from a detection method that is not very reliable can be modeled as a subnormal fuzzy set. This set would have a high core value (indicating severity) but a low height (indicating reliability).

In addition, \textit{ranking indices} offer a structured method to reconcile severity, confidence, and an organization's risk profile. These indices, rather than yielding a singular "best" ranking, enable an organization to adjust its alert prioritization strategy according to its risk appetite. Consequently, a more conservative organization might prioritize alerts with high confidence, whereas a risk-aggressive organization could choose to prioritize alerts with high severity.

\subsection{Contributions and Organization}

This paper makes four main contributions:

\begin{enumerate}
    \item {\textbf{Novel Application}}: We propose the first application of subnormal Gaussian fuzzy numbers to intrusion detection alert prioritization.
    
    \item {\textbf{Framework}}: We present an alert ranking framework, based on a clear understanding of risk and a parameterized ranking index. It explicitly considers the severity of alerts, the confidence in their detection, and the organization's risk tolerance.
    
    \item {\textbf{Empirical Validation}}: We validate the framework using real-world data from the CIC-IDS2017 dataset. Our results show that we see significant improvements in performance when the alert data is noisy or uncertain.
    
    \item {\textbf{Practical Insights}}: This research offers a detailed analysis of how key factors, such as detection confidence and risk attitude, affect alert prioritization. As a result, this research provides practical guidance for organizations using this system.
\end{enumerate}

The remainder of this paper is organized as follows: Section~\ref{sec:relatedWork} reviews related work on alert prioritization and fuzzy logic in cybersecurity; Section~\ref{sec:methodology} presents mathematical preliminaries, problem formulation, and the proposed framework; Section~\ref{sec:experiment} details the experimental setup; Section~\ref{sec:results} presents results and analysis; and Section~\ref{sec:conclusion} concludes with directions for future work.

\section{Related Work}
\label{sec:relatedWork}

IDSs have evolved significantly over the past two decades. Early IDS such as Snort and Suricata employed signature-based detection, matching network traffic with a database of known attack patterns~\cite{Roesch1999}. In recent decades, machine learning approaches have been applied to IDSs, using techniques such as Random Forests, neural networks, and Support Vector Machines (SVM) to detect anomalous behavior~\cite{Sommer2010, Lippmann2000, Breiman2001}. More recently, deep neural networks~\cite{Javaid2016}, recurrent neural networks for sequential attack detection~\cite{Staudemeyer2015}. These advances have improved detection accuracy, but have also increased the problem of alert volume --- machine learning-based IDS often generate more alerts than signature-based systems because they are more sensitive to deviations from normal behavior.

The alert fatigue problem has been extensively documented in the security operations literature. Julisch~\cite{Julisch2003} demonstrated that alert correlation can significantly reduce alert volume, but the fundamental problem of prioritization remains. In recent years, Alahmadi et al.~\cite{Alahmadi2022false} conducted a qualitative study that examined the perspectives of SOC analysts on security alarms, finding that alert validation is a tedious task that causes burnout. They identified five critical properties--Reliable, Explainable, Analytical, Contextual, Transferable--that are missing from current security tools. Jalalvand et al.~\cite{Jalalvand2024alert} provided a comprehensive survey of alert prioritization methods, noting that organizations receive more than $10,000$ alerts daily with 50\% FPs. In practice, alert prioritization is often performed manually by security analysts using domain knowledge and experience. Despite the operational importance of alert prioritization, this comprehensive systematic survey also confirms that there is limited research on this problem.

Recent work in machine learning-based security has emphasized the importance of uncertainty quantification. Bayesian deep learning approaches have been applied to IDS to provide confidence estimates for predictions~\cite{Wong2023}. Malhotra et al.~\cite{Malhotra2016} applied uncertainty quantification to anomaly detection in time series. Guo et al.~\cite{Guo2017} studied calibration of neural network predictions for security applications. Deep learning methods often treat uncertainty through probabilistic outputs or Bayesian approximations. A notable limitation of these probabilistic frameworks is that they require precise probability distributions, which may not be available in practice.

Traditional detection methods often struggle to accurately identify new or previously unknown attacks. A significant limitation of IDSs that use machine learning, particularly deep learning, is their lack of interpretability. Security analysts, who need to trust the system, often need to understand the reasons behind an alert and why it was given a specific priority. Ribeiro et al.~\cite{Ribeiro2016} developed LIME, a method to explain how machine learning models make predictions. Montavon et al.~\cite{Montavon2017} surveyed interpretability methods for deep learning. However, fuzzy logic offers inherent interpretability. That is, membership functions and ranking indices have clear linguistic interpretations, and analysts can understand the reasoning behind alert prioritization~\cite{Zimmermann2010}. Our fuzzy-based alert prioritization framework provides interpretable reasoning about alert priorities, complementing the detection capabilities of ML-based IDS.

CVSS is the standard metric for assessing vulnerability severity. Mell et al.~\cite{Mell2007} introduced CVSS v2, which has become widely adopted. The CVSS v3 specification~\cite{CVSS2015} improved on v2 with additional metrics. Despite its ubiquity, CVSS provides only a direct estimate of severity and does not account for organizational context or uncertainty. Allodi and Massacci~\cite{Allodi2012} studied the relationship between CVSS scores and the actual availability of exploits. Jacobs et al.~\cite{Jacobs2019} analyzed the limitations of CVSS for vulnerability prioritization. Our approach extends CVSS by incorporating the organizational context through contextual factors and modeling uncertainty through the spread parameter ($\sigma$) of the fuzzy number.

Julisch~\cite{Julisch2003} pioneered alert correlation using clustering techniques. Alert correlation groups related alerts into clusters to reduce alert volume, while our work focuses on ranking individual alerts by priority. Although complementary, correlation alone does not solve the prioritization problem, even after clustering, analysts must still decide which alerts or clusters to investigate first. Our framework provides fine-grained per-alert prioritization that can be applied after correlation.

Fuzzy logic has been used in various cybersecurity areas, including finding unusual activities, assessing threats, and detecting intrusions. Fuzzy logic can work with partial information and expert assessments, provides interpretable reasoning, and naturally represents the concept of partial membership and partial reliability~\cite{Zimmermann2010}. Tsang et al.~\cite{Tsang2007} proposed a genetic-fuzzy rule mining system for intrusion detection. More recently, Kar and Mishra~\cite{Kar2012} developed a fuzzy-based approach to detect HTTP-based attacks. Meng et al.~\cite{Meng2013} applied fuzzy logic to alert prioritization in network security, but their approach did not explicitly model detection confidence or use subnormal fuzzy sets. Existing applications of fuzzy logic to IDS typically do not explicitly address alert prioritization, nor do they model detection confidence or organizational risk attitude. Our framework based on subnormal Gaussian fuzzy numbers (SGFN) provides a more interpretable model of uncertainty. This allows security analysts to more easily understand and even tune the system's risk-attitude based on the three intuitive sources of uncertainty: severity, detection confidence, and classification ambiguity. Additionally, our framework naturally handles novel attacks by assigning low detection confidence, which can be adjusted based on organizational risk tolerance. It is complementary to the detection-focused approaches mentioned above. With this combination, organizations can achieve both high detection accuracy and effective alert management. To this end, we use a machine learning baseline IDS and focus on the main problem of ranking and prioritizing its alerts.

\section{Methodology}
\label{sec:methodology}

\subsection{Overview of the Proposed Framework}

\begin{figure}[t]
\centering
\includegraphics[width=\linewidth]{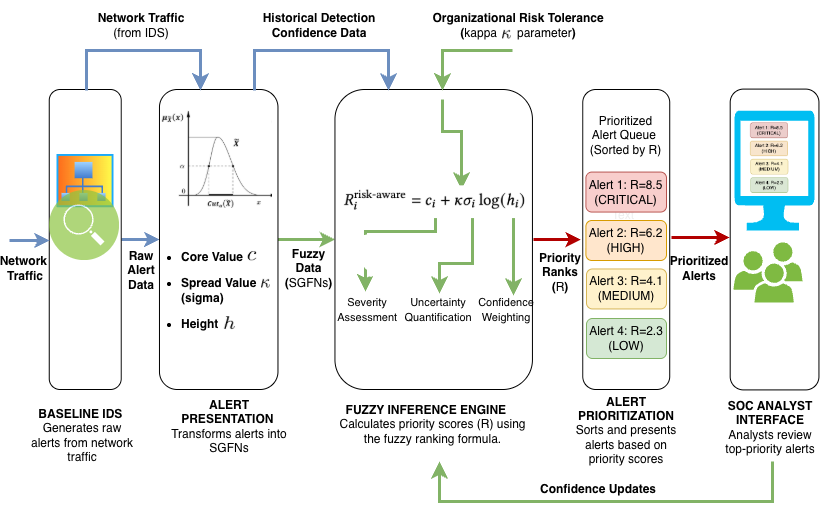}
\caption{The framework of the proposed intrusion detection alert prioritization system.}
\label{fig:Framework.png}
\end{figure}

We propose a comprehensive framework for IDS alert prioritization based on SGFNs and ranking indices. The framework has four steps: (1) represent alerts as fuzzy numbers, (2) extract and calibrate parameters, (3) calculate risk-averse rankings, and (4) rank alerts. Figure~\ref{fig:Framework.png} illustrates the overall architecture of the proposed system: Raw alerts from a baseline IDS are converted to SGFN with three parameters: core value $c$ (threat severity), spread value $\sigma$ (uncertainty in impact) and height value $h$ (detection confidence). The fuzzy ranking engine applies the ranking formula $R^{benefit} = c + \kappa \sigma  \log(h)$ to compute priority scores. The risk-attitude parameter $\kappa$, which we sweep over 0--2 (with the standard risk parameter $\kappa=1$ as the default in our main results), allows organizations to customize their alert prioritization based on their risk tolerance. After this, the alerts are sorted by their priority scores and presented to SOC analysts in a prioritized queue, which helps them investigate threats more effectively.

The primary advancement of our methodology lies in its explicit incorporation of three distinct sources of uncertainty within the alert prioritization process: uncertainty regarding threat severity, uncertainty regarding detection confidence, and the organization's risk attitude. In contrast to conventional methods, our framework offers a structured approach for evaluating alerts under uncertainty, enabling organizations to adjust their security posture according to their specific risk tolerance levels.

\subsection{Mathematical Preliminaries and Definitions}

\subsubsection{Subnormal Gaussian Fuzzy Numbers (SGFNs)}
Given that $p,q,r\in\mathbb{R}$, where $p<q<r$ and $h \in\left(0,1\right]$, an SGFN $\tilde{A}$ on $\mathbb{R}$ is a fuzzy subset characterized by a membership function $\mu_{\tilde{A}}: \mathbb{R} \to [0,1]$ satisfying the following properties~\cite{Chen1985, Zadeh1965}:

\begin{enumerate}
    \item $\mu_{\tilde{A}}$ is continuous
    \item $\mu_{\tilde{A}}(x) = 0$ for all $x \notin [p, r]$ (compact support)
    \item $\mu_{\tilde{A}}$ is strictly increasing on $[p, q]$ and strictly decreasing on $[q, r]$
    \item $\mu_{\tilde{A}}(q) = h$ where $h \in (0, 1]$ (height parameter)
\end{enumerate}

An SGFN $\tilde{A} = \langle (c_{\tilde{A}}, \sigma_{\tilde{A}}); h_{\tilde{A}} \rangle$ has a Gaussian membership function defined as:
\begin{equation}
\mu_{\tilde{A}}(x) = h_{\tilde{A}} \exp\left(-\frac{1}{2}\left(\frac{x - c_{\tilde{A}}}{\sigma_{\tilde{A}}}\right)^2\right)
\label{eq:sgfn_membership}
\end{equation}
where:
\begin{itemize}
    \item $c_{\tilde{A}} \in \mathbb{R}$ is the \textit{center} (core), representing the most likely value
    \item $\sigma_{\tilde{A}} > 0$ is the \textit{spread} (standard deviation), representing uncertainty or dispersion
    \item $h_{\tilde{A}} \in (0, 1]$ is the \textit{height}, representing the maximum membership degree (reliability)
\end{itemize}

The Gaussian membership function has several desirable properties: it is smooth, has a compact parametric representation, and supports closed-form arithmetic operations.

\subsubsection{$\alpha$-Cut Representation}
The $\alpha$-level set (or $\alpha$-cut) of an SGFN $\tilde{A}$ is the set of all elements with membership degree at least $\alpha$:
\begin{equation}
\begin{split}
\tilde{A}_{\alpha} &= \{x \in \mathbb{R} \mid \mu_{\tilde{A}}(x) \geq \alpha, \alpha \in [0, h_{\tilde{A}}]\} \\
&= [\tilde{A}_L(\alpha), \tilde{A}_R(\alpha)]
\label{eq:alpha_cut}
\end{split}
\end{equation}
For an SGFN with Gaussian membership function, the left and right endpoints of the $\alpha$-cut are:
\begin{equation}
\tilde{A}_L(\alpha) = c_{\tilde{A}} - \sigma_{\tilde{A}} \sqrt{-2 \ln(\alpha / h_{\tilde{A}})}
\label{eq:alpha_left}
\end{equation}
\begin{equation}
\tilde{A}_R(\alpha) = c_{\tilde{A}} + \sigma_{\tilde{A}} \sqrt{-2 \ln(\alpha / h_{\tilde{A}})}
\label{eq:alpha_right}
\end{equation}
The $\alpha$-cut representation is useful for understanding the range of possible values at different levels of confidence.

\subsubsection{Ranking Index for Fuzzy Numbers}

For a benefit-type fuzzy number (where higher values are preferable), we use the ranking index proposed in our prior work~\cite{Akdemir2025}. 

\begin{equation}
R^{\text{benefit}}(\tilde{A}; \kappa) = \underbrace{c_{\tilde{A}}}_{\text{base severity}} + \underbrace{\kappa \sigma_{\tilde{A}} \log(h_{\tilde{A}})}_{\text{risk adjustment}}
\label{eq:ranking_decomposition}
\end{equation}
where $\kappa \geq 0$ is the \textit{risk-attitude parameter}.

The first term is the base severity (core value). The second term is a risk adjustment that penalizes alerts with high uncertainty ($\sigma_{\tilde{A}}$) or with low confidence ($h_{\tilde{A}}$) since the value log($h_{\tilde{A}}$) is negative (assuming base-10 logarithm). This penalty is also controlled by the risk-attitude parameter $\kappa$.

\subsection{Alert Representation as SGFNs}

The fundamental advancement of our methodology is the representation of each IDS warning as an SGFN. This representation captures three key aspects of alert assessment: severity, uncertainty, and detection confidence.

\subsubsection{Core Value $c_i$ (Threat Severity)}

The core value $c_i$ of the alert $A_i$ represents the threat severity score. We compute this using the CVSS score of the corresponding attack type and organizational context (contextual factor $\text{cf}_i$):
\begin{equation}
c_i = \text{CVSS}(\text{attack\_type}_i) \times \text{cf}_i
\label{eq:core_value}
\end{equation}
where:
\begin{itemize}
    \item $\text{CVSS}(\text{attack\_type}_i) \in [0, 10]$ is the CVSS base score for the detected attack type, obtained from the National Vulnerability Database (NVD) or threat intelligence sources
    \item $\text{cf}_i \in [0, 1]$ is an organizational context factor that adjusts severity, based on whether the target system is critical
\end{itemize}

CVSS provides a numerical score, based on factors such as attack vector, required privileges, user interaction, and potential impact. Security teams prioritize vulnerabilities according to CVSS scores for attack types. For example, a remote SQL injection vulnerability that requires no authentication and allows full database compromise might receive a CVSS score of 9.8 (Critical) due to its significant impact on confidentiality, integrity, and availability (CIA). A stored cross-site scripting (XSS) vulnerability that necessitates user interaction but facilitates session hijacking could be assessed at approximately 7.2, indicating a high severity. Conversely, a reflected XSS vulnerability, characterized by a restricted impact and the requirement of user interaction, might receive a medium rating of 5.4. Furthermore, a local denial-of-service vulnerability, which exclusively compromises availability, could be assigned a low score of 3.1.

The contextual factor $\text{cf}_i$ is the environment-specific conditions that influence how severe or risky a vulnerability is for a particular organization, beyond its generic technical severity. Thus, it means that the same vulnerability can have different risk levels for different organizations. Organizations generally use tools to calculate these scores with respect to their modified impact metrics on CIA and to weigh them according to their security requirements. Throughout the paper, we consider the values of $\text{cf}_i$ as the following:
\begin{equation}
\text{cf}_i = \begin{cases}
1.0 & \text{if target is critical infrastructure} \\
0.8 & \text{if target is important business system} \\
0.5 & \text{if target is non-critical system} \\
0.2 & \text{if target is isolated test system}
\end{cases}
\label{eq:contextual_factor}
\end{equation}
This enables the framework to adapt to organizational priorities. For example, an attack on a critical system is weighted more heavily than the same attack on a non-critical system.
In our experiments, we assign a deterministic per-alert contextual factor $\text{cf}_i \in [0.2, 1.0]$ using a hash of the alert id and attack type, then map it to the nearest categorization. This increases contextual variability without introducing randomness.

\subsubsection{Spread Value $\sigma_i$ (Uncertainty in Severity)}

The spread value $\sigma_i$ represents uncertainty in severity assessment. We compute this as a percentage of the core value by multiplying it by the uncertainty factor $\text{uf}_i$. Its value depends on the attack types and reflects our confidence in the severity assessment. Well-known attacks have well-understood impacts, and thus low uncertainty. Novel attacks have unknown impacts and thus high uncertainty.
\begin{equation}
\sigma_i = c_i \times \text{uf}_i (\text{attack\_type}_i)
\label{eq:spread_value}
\end{equation}
where $\text{uf}_i (\text{attack\_type}_i) \in [0, 0.5]$ depends on the attack class $i$ and $\sigma_i$ is the standard deviation derived by multiplying the uncertainty factor and severity value. These values are assigned based on how well-known the attack is. For instance, values may be 0.1 for well-known attacks (e.g. SQL injection, buffer overflow, brute force, port scan), 0.2 for moderately known attacks (e.g., specific DoS and DDoS variants, malware), 0.35 for emerging and poorly known threats (e.g., new malware variants, anomalous traffic), and 0.5 for novel or zero-day attacks. Although this approach to calculate $\sigma_i$ requires defining the value \text{uf} for each attack type, it is proportional to severity: higher severity means higher uncertainty.~\footnote{For each attack type, corporations need to validate their calculated uncertainty scores to see whether it matches with their experience and adjust accordingly if necessary as in Definition~\ref{eq:uncertainty_factor}.}
\begin{equation}
\text{uf}_i (\text{attack\_type}_i) = \begin{cases}
0.1 & \text{if it is well-known} \\
0.2 & \text{if it is moderately known} \\
0.35 & \text{if it is emerging} \\
0.5 & \text{if it is novel/zero-day}
\end{cases}
\label{eq:uncertainty_factor}
\end{equation}
\subsubsection{Height Value $h_i$ (Detection Confidence)}
The height value $h_i \in [0, 1.0]$ encodes the detector reliability for this alert (the height of the fuzzy number in the SGFN representation), not the raw confidence score $p_i$ from the baseline IDS. We first compute class-level $h_{\text{class}}$ from validation performance (Equation~\ref{eq:height_class}). the height parameter reflects the actual historical reliability of detecting that attack class. Hence, the type of detection method that generated the alert is substantial. For instance, if the detection method is reliable with few FPs (signature-based methods such as Snort rules) then the height value should be high; If the detection method is moderately reliable with many FPs (heuristic based, machine learning, and anomaly detection) then the height value should be around median; Lastly, if the detection method is unreliable with too many FPs (novel behavior) then the height value should be low. However, before determining instance‑level detection confidence $h_i$, we first compute the class-level height value $h_{\text{class}}$ for each attack type depending on the historical performance of the detection method during IDS training on a validation dataset as in Equation~\ref{eq:height_class}. 
\begin{equation}
h_{\text{class}} = \operatorname{clip}\left(0.5 + \alpha(\text{F1}_i - 0.5), h_{\min}, h_{\max}\right)
\label{eq:height_class}
\end{equation}
We clamp $0.5 + \alpha(\text{F1}_i - 0.5)$ to $[h_{\min}, h_{\max}]$ to avoid saturation and use $\alpha=0.9$, $h_{\min}=0.05$, $h_{\max}=0.95$ in our experiments. $\alpha$ controls shrinkage strength (1.0 means no shrinkage; smaller values apply stronger smoothing) and reduces variance in small or imbalanced classes. Appendix~\ref{app:sensitivity} reports a sensitivity analysis for $\alpha$, $h_{\min}$, $h_{\max}$, and a global uncertainty scaling factor, showing that the ranking results are stable within reasonable ranges.

To determine $h_i$ we retain instance-level variability by capping the class-level height with the per-alert model confidence $p_i$ as in Equation~\ref{eq:height_instance}. Using the min implements a conservative cap: an individual alert cannot be more reliable than the class has proven to be, and a low per-alert confidence should down-weight even a strong class. It is conservative in a principled way: the system refuses to trust an alert beyond what historical performance and current evidence jointly justify.
\begin{equation}
h_i = \min(h_{\text{class}}, p_i)
\label{eq:height_instance}
\end{equation}
where $p_i = P(\text{attack}\mid x_i)$ denotes the IDS's raw probability for alert $i$, showing how certain this alert is an attack. This may be different for each alert of the same class and varies unpredictably.

$\text{Precision}_i$ is the precision of the baseline IDS in this attack class (fraction of alerts that are true positives). Although it requires manual verification and historical data, this is the most accurate method and reflects actual IDS performance. Here, $i$ indexes the attack class, $TP_i$ is the number of true positives, $FP_i$ is false positives, and $FN_i$ is false negatives.
\begin{equation}
\text{Precision}_i = \frac{TP_i}{TP_i + FP_i}
\label{eq:precision}
\end{equation}

$\text{Recall}_i$ also accounts for false negatives of the baseline IDS.
\begin{equation}
\text{Recall}_i = \frac{TP_i}{TP_i + FN_i}
\label{eq:recall}
\end{equation}
$\text{F1}_i$ is the F1-score of the baseline IDS in this attack class (harmonic mean of precision and recall). This is particularly important for detection quality because precision alone ignores missed detections, and hence F1 balances FPs and FNs:
\begin{equation}
\text{F1}_i = 2 \times \frac{\text{Precision}_i \times \text{Recall}_i}{\text{Precision}_i + \text{Recall}_i}
\label{eq:f1}
\end{equation}
We include precision and recall for diagnostics, but we use F1 to derive the class height using Equation~\ref{eq:height_class}.

For attack classes not seen during validation (novel attacks), we use a conservative estimate $h_{\text{class}} = 0.5$ and then apply the same capping as $p_i$. This reflects the fact that we have no empirical evidence of the reliability of the detection method on novel attacks.

The key insight is that the height parameter captures the \textit{reliability of the detection method}, not the raw confidence score from the IDS. This is crucial because

\begin{enumerate}
    \item a high-confidence alert from an unreliable detection method should be de-prioritized
    \item a low-confidence alert from a reliable detection method should be prioritized
    \item the height parameter is attack-class specific, allowing different reliability for different attack classes
\end{enumerate}

To illustrate the representation of alerts as SGFNs, assume a raw alert from an ML-based anomaly detector with attack type SQL Injection, CVSS score 7.5, and target customer database. The calculations for the fuzzy number $\langle (c, \sigma); h) \rangle$ are:

\begin{itemize}
\item {\textbf{Contextual Factor}}: 0.8 (important target), hence $c = 7.5 \times 0.8 = 6.00$

\item {\textbf{Uncertainty Factor}}: 0.1 (well-known attack), hence $\sigma = 6.00 \times 0.15 = 0.90$

\item {\textbf{Height Value}}: Let us assume historical data as: 45 TP, 35 FP, 15 FN, so $\text{Precision} = 45/(45+35) = 0.56$ and $\text{Recall} = 45/(45+15) = 0.75$, $\text{F1}= 2 \times (0.56 \times 0.75) / (0.56 + 0.75) = 0.64$. With $\alpha=0.9$, $h_{\text{class}} = 0.5 + 0.9(0.64 - 0.5) = 0.626$ (clipped to $[0.05, 0.95]$); the height of the instance is $h_i=\min(h_{\text{class}}, p_i)$.
\end{itemize}

\subsection{Alert Prioritization Algorithm}

Given a set of alerts $\mathcal{A} = \{A_1, A_2, \ldots, A_n\}$ and a risk-attitude parameter $\kappa$, we first convert each raw alert into an SGFN. Second, we calculate the priority score using the ranking formula in Equation~\ref{eq:ranking_decomposition}. Here, the ranking score is a single numerical value that combines severity, uncertainty, and detection confidence into a unified prioritization metric. Third, we sort the alerts by priority score in descending order so that the alerts with higher ranking scores have higher priority. Finally, we present ranked alerts to analysts with interpretable information. More details of each step of the alert prioritization algorithm are given in Algorithm~\ref{alg:alert_ranking}. 

\begin{algorithm}[t]
\caption{Context-Aware Alert Representation and Ranking}
\label{alg:alert_ranking}
\begin{algorithmic}[1]
\REQUIRE Alert set $\mathcal{A} = \{A_1, A_2, \dots, A_N\}$, uncertainty factors, CVSS scores, target criticality to derive the contextual factor, detection metrics, height smoothing parameters $(\alpha, h_{\min}, h_{\max})$, parameter $\kappa$
\ENSURE Ranked list of alerts

\STATE \textbf{Step 1: Alert Representation}
\FOR{each alert $A_i \in \mathcal{A}$}
    \STATE Determine attack type $\text{attack\_type}_i$
    \STATE Retrieve CVSS base score $\text{CVSS}_i \gets \text{CVSS}(\text{attack\_type}_i)$
    \STATE Compute contextual factor $\text{cf}_i \gets \text{cf}_i(A_i)$
    \STATE Compute core severity value $c_i \gets \text{CVSS}_i \times \text{cf}_i$
    \STATE Compute spread value $\sigma_i \gets c_i \times \text{uf}_i(\text{attack\_type}_i)$
    \IF{attack type is known}
    \STATE Set class confidence $h_{\text{class}} \gets \operatorname{clip}(0.5 + \alpha(\text{F1}(\text{attack\_type}_i) - 0.5), h_{\min}, h_{\max})$
    \STATE Set detection confidence $h_i \gets \min(h_{\text{class}}, p_i)$

    \ELSE
        \STATE Set detection confidence $h_i \gets \min(0.5, p_i)$
    \ENDIF
    \STATE Construct SGFN alert representation $\tilde{A}_i \gets \langle (c_i, \sigma_i); h_i \rangle$
\ENDFOR

\STATE \textbf{Step 2: Ranking Score Computation}
\FOR{each alert representation $\tilde{A}_i$}
    \STATE Compute ranking score
    \[
    R^{\text{risk-averse}}_i \gets c_i + \kappa \sigma_i  \log(h_i)
    \]
\ENDFOR

\STATE \textbf{Step 3: Alert Sorting}
\STATE Sort alerts in descending order by ranking score $R^{\text{risk-averse}}_i$

\STATE \textbf{Step 4: Alert Presentation}
\FOR{each alert in sorted order}
    \STATE Present attack type, target system, $c_i$, $\sigma_i$, $h_i$, and $R^{\text{risk-averse}}_i$
\ENDFOR

\RETURN Ranked alert list
\end{algorithmic}
\end{algorithm}

About the computational complexity of the alert prioritization algorithm: Given that $n$ is the number of alerts, the complexity of the representation of alerts and ranking score computation is linear $O(n)$. Moreover, sorting in terms of ranking scores is $O(n \log n)$. The overall complexity is $O(n \log n)$, which is efficient even for large numbers of alerts.

\subsection{Interpretation of the Ranking Score}
The ranking score computed for each alert $R^{\text{benefit}}_i(\tilde{A}_i; \kappa) = c_i + \kappa \sigma_i \log(h_i)$, we call it $R_i^{\text{risk-averse}}$, has a clear interpretation:

\textbf{Base Severity Term ($c_i$):} This is the main driver of alert priority. Higher severity alerts are prioritized.

\textbf{Risk-Attitude Parameter ($\kappa$):} This parameter allows organizations to tune their security posture. Table~\ref{tab:risk_profiles} describes what different $\kappa$ values mean.
\begin{table}[h]
\caption{Risk Aversion Profiles based on $\kappa$ values}
\label{tab:risk_profiles}
\begin{tabular}{p{1.4cm} p{3.2cm} p{8.0cm}}
\toprule
\textbf{Value of $\kappa$} & \textbf{Risk Profile} & \textbf{Description} \\ \midrule
$\kappa = 0.0$ & Pure Severity-based & Aggressive; focuses on high severity regardless of confidence. \\ \hline
$\kappa = 0.5$ & Moderate Risk Aversion & Balanced approach between severity and confidence. \\ \hline
$\kappa = 1.0$ & Standard Risk Aversion & Default; balances severity and confidence equally. \\ \hline
$\kappa > 1.0$ & High Risk Aversion & Conservative; prioritizes high-confidence alerts over severity. \\ \bottomrule
\end{tabular}
\end{table}

\textbf{Risk Adjustment Term ($\kappa \sigma_i \log(h_i)$):} This term adjusts the priority based on the risk-attitude parameter, uncertainty, and confidence. Since $h_i \in (0, 1]$, we have $\log(h_i) \leq 0$, hence the lower confidence value $h_i$ means the bigger negative number, which penalizes more. As $\kappa \sigma_i \log(h_i) \leq 0$, meaning that the adjustment term is negative; higher risk-aversion, higher uncertainty, and lower confidence reduce the priority of alerts with higher penalties.

\subsection{Handling Edge Cases and Calibration}

\subsubsection{Novel or Zero-Day Attacks}

For attacks not seen during validation, we cannot compute the detection confidence from historical data. We handle this by assigning a conservative class height: $h_{\text{class}} = 0.5$ and applying $h_i = \min(h_{\text{class}}, p_i)$

\subsubsection{Missing or Incomplete Information}

If some information is missing (e.g., CVSS score not available), the default values shown in Table~\ref{tab:default_values} can be used:
\begin{table}[h]
\caption{Default values for missing vulnerability metrics}
\label{tab:default_values}
\begin{tabular}{lcl}
\toprule
\textbf{Missing Parameter} & \textbf{Default Value} & \textbf{Interpretation} \\ \midrule
CVSS Score ($\text{CVSS}_i$) & $5.0$ & Medium severity \\
Contextual Factor ($\text{cf}_i$) & $0.5$ & Moderate importance \\
Detection Confidence ($h_i$) & $0.5$ & Moderate reliability \\
IDS Attack Probability ($p_i$) & $0.5$ & Default per-alert attack probability \\ \bottomrule
\end{tabular}
\end{table}

\subsubsection{Calibrating the Risk-Attitude Parameter ($\kappa$)}

The risk-attitude parameter $\kappa$ should be calibrated based on organizational risk tolerance. Thus, an organization would start with $\kappa = 1$ (default, balanced) and analyze the top-ranked alerts over a period of time. If too many FPs are in the top-ranked alerts, the analyst can increase $\kappa$ (more conservative) and if critical attacks are being missed, s/he can decrease it (more aggressive). This can be repeated until the ranking aligns with expectations.

\subsubsection{Calibrating Detection Confidence ($h_i$)}

The detection confidence should be computed from a validation data set. Thus, we first train the baseline IDS in the training set, then evaluate in the validation set (15-20\% of the data). Next, we compute precision, recall, and F1 for each attack class and derive $h_{\text{class}}$ from the smoothed F1 (Equation~\ref{eq:height_class}), then cap with $p_i$ using Equation~\ref{eq:height_instance}. For novel attacks, we start from $h_{\text{class}}=0.5$ (smoothed) and update as more data become available.

\subsection{Integration with Existing IDS}

Our framework is designed to work with any baseline IDS (signature-based, ML-based, or hybrid). During the integration process, the baseline IDS generates alerts with the attack type and confidence score. Our framework reads these alerts and extracts the attack type. For each alert, we look up the CVSS score, contextual factor, uncertainty factor, and detection confidence, compute the ranking score, and present ranked alerts to the analyst. The framework can be deployed as a post-processing module on top of existing IDS. This modular design allows organizations to adopt our framework without replacing their existing IDS infrastructure.

\subsection{Comparison with Baseline Ranking Approaches}

Our framework combines severity, uncertainty, and confidence with our underlying ranking as in Equation~\ref{eq:ranking_decomposition}. This approach is theoretically sound, allows for adjustments to risk preferences, and explicitly considers uncertainty. To better understand our method, we compare it with three standard ranking techniques:

\begin{enumerate}
    \item \textbf{Baseline 1: Severity-Only Ranking}: Ranks alerts purely by CVSS score: $R_i^{\text{severity}} = c_i$. This ignores detection confidence and leads to wasted analyst time investigating low-confidence, high-severity alerts.

    \item \textbf{Baseline 2: Confidence-Only Ranking}: Uses only confidence for ranking, sorts alerts based solely on the model's confidence for each alert: $R_i^{\text{confidence}} = p_i$. This method does not consider severity, which could lead to missing important attacks that are harder to detect accurately.

    \item \textbf{Baseline 3: Weighted Sum Ranking}: Combines severity and per-alert confidence with ad-hoc weights: $R_i^{\text{weighted}} = 0.5 \text{norm}(c_i) + 0.5 \text{norm}(p_i)$. This does not account for the organizational risk attitude or uncertainty.
    
\end{enumerate}

\section{Experimental Setup}
\label{sec:experiment}

\subsection{Dataset}

We evaluate our framework using the CIC-IDS2017 dataset~\cite{Sharafaldin2018}. The dataset is a publicly available, modern, comprehensive intrusion detection dataset created by the Canadian Institute for Cybersecurity. It has been widely used in the literature, allowing us to compare our framework with other related works. It contains real-world network traffic captured over five days in a controlled environment. The dataset is ideal for our evaluation because it contains diverse, realistic attack types (both common attacks and emerging threats), and provides fine-grained flow-level labels.

The data set consists of 2,830,743 total network flows, collected in 5 continuous day network traffic, 80.3\% of which are benign (2,273,643) and 19.7\% of which are attack flows (557,100). It also covers 78 network flow features (source/destination IP, ports, protocol, packet statistics, etc.) and 11 attack types (mapped to 8 classes); Table~\ref{tab:cyber_attack_vectors} shows the condensed categories.

\begin{table}[h]
\caption{Classification of Common Cyber Attack Types}
\label{tab:cyber_attack_vectors}
\begin{tabular}{p{2.8cm} p{3.2cm} p{7.6cm}}
\toprule
\textbf{Category} & \textbf{Attack Type} & \textbf{Description} \\
\midrule
Credential Guessing / Password Attacks
& Brute Force (FTP-Patator, SSH-Patator)
& Repeated, systematic attempts to guess login credentials for network services such as FTP and SSH. \\
\midrule
System Application Attacks
& Web Attacks (Brute Force, XSS, SQL Injection)
& Exploitation of web application vulnerabilities, including brute-force login attempts and injection attacks. \\
\midrule
Vulnerability Exploitation
& Heartbleed
& Exploitation of a cryptographic vulnerability in OpenSSL that allows attackers to read sensitive memory contents. \\
\midrule
Denial of Service
& DoS (Hulk, GoldenEye, Slowloris, Slowhttptest)
& Resource-exhaustion attacks that degrade or deny service availability. \\
\midrule
Distributed Denial of Service
& DDoS
& Coordinated attacks that overwhelm a target with excessive traffic from multiple sources. \\
\midrule
Reconnaissance
& Port Scan
& Probing network ports to discover open services and potential weaknesses. \\
\midrule
Malware / Botnets
& Bot
& Botnet-controlled traffic used for spam, scanning, or coordinated attacks. \\
\midrule
Lateral Movement / Intrusion
& Infiltration
& Unauthorized internal access or data exfiltration after initial compromise. \\
\bottomrule
\end{tabular}
\end{table}

For data preparation and splitting, we follow standard practice for IDS evaluation:
\begin{itemize}
    \item \textbf{Feature Normalization}: We normalize all 78 features to the range [0, 1] using min-max scaling to ensure that no single feature dominates due to scale differences.
    \item \textbf{Train/Validation/Test Split}: Sort day files by weekday. If the day-based train split (Mon/Tue) has at least 5\% attacks, use Mon/Tue for training, Wed for validation, Thu/Fri for test.
    \item \textbf{Attack Type Mapping}: Map the 11 attack types to 8 attack classes by grouping variants that represent the same attack family to consolidate many raw labels into a smaller and consistent set of attack classes 
    \item \textbf{Imbalance Handling}: Keep the natural imbalance to avoid the model learning "everything is benign" when attacks are rare (below 5\%) unless the fallback is triggered to a stratified split (50/20/30) to preserve learnability.
\end{itemize}

\subsection{Baseline IDS}

To evaluate our alert prioritization framework, we use a machine learning-based baseline IDS Logistic Regression (LR) to generate alerts. The baseline IDS is trained to detect attacks and is used to produce confidence scores, which are later used to rank alerts in our prioritization framework. 

To this end, we employ Logistic Regression (LR) as the baseline IDS because it provides well-calibrated probabilities suitable for ranking analysis, offers interpretability through linear coefficients, and efficiently handles class imbalance. Importantly, when detectors are near-perfect (such as Random Forest), different ranking methods converge, making comparisons uninformative. A moderately-performing detector, such as LR, better demonstrates the value of ranking under uncertainty. The LR hyper-parameters are solver \texttt{saga}, $C=1.0$, maximum iterations $=1000$ and class weights set to balanced. We apply sigmoid probability calibration on the validation set and use the calibrated class probabilities as per-alert confidence $p_i$. We also evaluate auxiliary detectors, Random Forest and Isolation Forest, in Appendix~\ref{subsec:aux-detectors} to check detector sensitivity.

For each attack class in the validation data set, we compute detection confidence metrics (precision, recall, and F1) and derive the class-level height $h_{\text{class}}$ using the smoothed F1 formulation in Equation~\ref{eq:height_class}. During ranking, we use $h_i = \min(h_{\text{class}}, p_i)$ to cap the class reliability by the per-alert confidence.

To stress the detector without changing the ranking logic, we also train a reduced-feature LR using only TCP-flag-related attributes (flags-only). This produces a deliberately noisier alert stream while keeping the ranking pipeline identical, enabling clearer comparison of ranking methods under realistic uncertainty.

\subsection{Performance Metrics}

To measure the quality of the ranking while accounting for the position of items in the ranked list, we use Normalized Discounted Cumulative Gain (NDCG)~\cite{JarvelinKekalainen2002}. Unlike binary metrics (true vs. false), NDCG can account for severity levels. It rewards systems more for placing "Critical" alerts at the top than "Medium" ones. NDCG ranges from 0 to 1, where 1 represents a perfect ranking (the higher, the better). It is particularly useful because it accounts for ranking position: finding a TP at position 1 is much better than finding it at position 100. NDCG$_{\text{rel}}$@k is a severity-adjusted variant. Using NDCG$_{rel}@k$, we directly measure what matters operationally: whether our ranking method successfully brings the most important, true attacks to the top of the alert queue.

 
\begin{equation}
\text{DCG@k} = \sum_{i=1}^{k} \frac{2^{rel_i} - 1}{\log_2(i+1)}
\label{eq:dcg}
\end{equation}
\begin{equation}
\text{NDCG@k} = \frac{\text{DCG@k}}{\text{IDCG@k}}
\label{eq:ndcg}
\end{equation}

where $rel_i$ is the relevance of the item at position $i$ (1 for TP 0 for FP), and IDCG@k is the DCG of the hypothetical perfect or ideal ranking (all TPs first). To emphasize severity while accounting for uncertainty, we adjust the original definition of relevance by severity: for TPs $c_i(1-\text{uf}_i)$ and 0 for benign alerts.

We denote the resulting metric by NDCG$_{\text{rel}}$@k and present multiple cutoffs. NDCG$_{\text{rel}}$@10 is the primary operational metric, reflecting the author's concept of an "impatient user" who only looks at the top results~\cite{JarvelinKekalainen2002}. Reasons to choose this metric: (1) most SOC analysts review 5 to 15 alerts per session, making the top 10 list the operational focus; (2) this cutoff best reveals where ranking decisions matter most—at the boundary between high-severity and borderline alerts; (3) larger $k$ values compress differences because the predicted-alert queue is attack-dominant, so the top 100 are nearly always true attacks. Moreover, we make use of NDCG$_{\text{rel}}$@50 and @100 to assess robustness beyond the top list, and NDCG$_{\text{rel}}$@500 for bootstrap significance testing. This multi-faceted approach allows for an evaluation of both the immediate usefulness of the ranking methods in practice and their overall stability.

We also evaluate the performance of the rankings within the confidence bands of the IDS probability $p_i$. We partition the alerts into three bins (0.3--0.5, 0.5--0.7, 0.7--1.0) and compute NDCG$_{\text{rel}}$@k on the band-restricted rankings. This highlights the differences in ranking methods when the detector's certainty is low.

In addition, we evaluate the sensitivity of the height-smoothing and uncertainty parameters by sweeping $\alpha$, $h_{\min}$, $h_{\max}$, and a global scaling of \text{uf}$_i$ in the full-feature LR setting. Appendix~\ref{app:sensitivity} shows NDCG$_{\text{rel}}$@10$_{\text{pred}}$ and NDCG$_{\text{rel}}$@100$_{\text{pred}}$ across these ranges, presenting only minor variation, indicating that the ranking behavior is stable within reasonable limits.

\subsection{Baseline Ranking Approaches}

We compare our ranking method with three baseline ranking approaches that represent different levels of sophistication in alert prioritization.

\subsubsection{Baseline 1: Severity-Only Ranking} This method ranks alerts solely on the basis of CVSS-based severity $c_i \in [0, 10]$. It is simple and transparent, but ignores detection confidence and can elevate low-confidence, high-severity alerts.
\begin{equation}
R_i^{\text{severity}} = c_i
\end{equation}
\subsubsection{Baseline 2: Confidence-Only Ranking} This method ranks alerts purely by the raw confidence score $p_i \in [0, 1]$ from the IDS. It prioritizes alerts the detector is most confident about, which can reduce FPs, but ignores severity and can miss critical attacks that are harder to detect. For unsupervised detectors (Isolation Forest in our experiments), anomaly scores are not calibrated probabilities, so Confidence-Only is a diagnostic baseline rather than a recommended method.
\begin{equation}
R_i^{\text{confidence}} = p_i
\end{equation}
\subsubsection{Baseline 3: Weighted Sum Ranking} This method combines severity and detector confidence with fixed weights (equal weights by default). It provides a middle ground between Baseline 1 and 2, but the weights are ad hoc and do not reflect the organizational risk attitude.

To ensure a fair comparison between severity and confidence components, we apply scale normalization as in Equation~\ref{eq:weighted}. This normalization ensures that both severity and confidence contribute meaningfully to the ranking score, preventing one component from dominating due to scale mismatch (CVSS severity typically ranges [5, 10] with variance $\sim$1.5, while detector confidence ranges [0.5, 0.95] with variance $\sim$0.03 for predicted alerts). 

\begin{equation}
\label{eq:weighted}
R_i^{\text{weighted}} = 0.5 \text{norm}(c_i) + 0.5 \text{norm}(p_i)
\end{equation}

where $\text{norm}(c_i) = \frac{c_i - min(c_i)}{max(c_i) - min(c_i)} $ normalizes the CVSS severity range and $\text{norm}(p_i)$ is also calculated in the same way, normalizing $p_i$ values. 

\subsubsection{Proposed Method: Risk-Averse Ranking} In our method, we rank alerts using the ranking index as in~\ref{eq:ranking_decomposition}. It explicitly models uncertainty and allows organizations to tune their risk attitude. Similarly, with this method, higher score equals to higher priority. Thus, it favors alerts with high severity and high confidence, and down‑weights alerts with high uncertainty or low confidence, in a controllable way. Although the method requires historical detection data to compute height values and requires parameter calibration, it explicitly models uncertainty in severity assessment, allowing tuning of risk attitude through parameter $\kappa$, and provides interpretable reasoning about alert priorities.

\section{Results and Analysis}
\label{sec:results}

\subsection{Baseline IDS Performance}

We evaluate the baseline IDS using standard classification metrics in the CIC-IDS2017 validation split to show the alert generator is competent. Additionally, we include a stressed LR trained in a reduced feature subset (flags-only) to create a more noisy alert stream for the analysis, presented in~\ref{sec:stress_test}. Table~\ref{tab:baseline_ids_metrics} shows the IDS performance for the two different environments.
\begin{table}[h]
\caption{Baseline IDS validation performance on CIC-IDS2017 (full vs.\ flags-only LR)}
\label{tab:baseline_ids_metrics}
\begin{tabular}{lcccc}
\toprule
\textbf{Features} & \textbf{Accuracy} & \textbf{Precision} & \textbf{Recall} & \textbf{F1-score} \\
\midrule
Full (all features) & 0.9268 & 0.8382 & 0.7789 & 0.8075 \\
Flags-only (stress) & 0.8089 & 0.5822 & 0.1055 & 0.1786 \\
\bottomrule
\end{tabular}
\end{table}

\subsection{Ranking Performance}
We first evaluate the ranking performance on the full-feature LR detector, presented in Table \ref{tab:precision_recall}. This scenario establishes a baseline understanding of method performance under favorable conditions. Our evaluation focuses on predicted-attack alerts (true attacks alerted by the IDS) to isolate the analyst's review queue and to avoid benign dominance in the metrics.
\begin{table}[h]
\caption{Predicted-alert ranking performance on CIC-IDS2017 (full-feature LR)}
\label{tab:precision_recall}
\begin{tabular}{p{2.8cm} p{2.4cm} p{2.4cm}p{2.4cm}p{2.4cm}} 
\toprule
\textbf{Method} & \textbf{NDCG$_{\text{rel}}$@10$_{\text{pred}}$} & \textbf{NDCG$_{\text{rel}}$@50$_{\text{pred}}$} & \textbf{NDCG$_{\text{rel}}$@100$_{\text{pred}}$} & \textbf{NDCG$_{\text{rel}}$@500$_{\text{pred}}$} \\
\midrule
Severity-Only & 0.9534 & 0.9836 & 0.9899 & 0.9970 \\
Confidence-Only & 0.5858 & 0.5858 & 0.5858 & 0.5858 \\
Weighted Sum & 0.9534 & 0.9836 & 0.9899 & 0.9970 \\
\textbf{Risk-Averse ($\kappa=1$)} & \textbf{1.0000} & \textbf{1.0000} & \textbf{1.0000} & \textbf{0.9974} \\
\bottomrule
\end{tabular}
\end{table}

With the full-feature detector, differences are intentionally small: Risk-Averse reaches NDCG$_{\text{rel}}$@10 = 1.0000 (+0.0466 over Severity-Only and Weighted Sum, +0.4142 over Confidence-Only), illustrating near-parity under strong IDS. A key finding in this scenario is that the Risk-Averse method with $\kappa=1$ achieves a perfect NDCG$_{\text{rel}}$@10 score of 1.0000. This perfect score is not a trivial result; it signifies that the method successfully ranked the top-10 most critical and true attacks in the exact order of their severity, without errors. This performance is attributed to the use of high-quality confidence scores ($h_i$) provided by the full-feature detector. The risk-attitude parameter ($\kappa=1$) provides a balanced default boost to high-severity alerts that also have high confidence, effectively breaking ties and correcting minor mis-orderings that a purely severity-based ranking might produce. For example, if two critical alerts have identical severity, the one with higher detection confidence is correctly promoted. This shows that even when a detector is strong, explicitly modeling the relationship between severity and confidence provides a more refined and operationally perfect ranking. Although the numerical gain over the baseline methods is modest in this high-signal environment, the perfect score validates the core principle of our method.

In this split, predicted alerts are overwhelmingly attacks; therefore, we emphasize NDCG$_{\text{rel}}$ across multiple $k$ cutoffs to separate ranking quality. The overall vs.\ predicted NDCG$_{\text{rel}}$ values differ only for Severity-Only, reflecting benign dominance in the full alert stream. In particular, the Severity-Only and Weighted-Sum methods produce identical scores. This occurs because the high-performance detector assigns uniformly high confidence scores to alerts in the predicted-attack queue, causing the confidence term in the weighted sum to have negligible impact on the final rank ordering, which remains dominated by severity.

As a limitation, the strong full-feature detector and attack-dominant queue compress ranking differences, so the stress test, confidence-band analysis, and statistical significance assessed in the next subsections provide more discriminative evidence.

\subsection{Stress-Test: Reduced-Feature LR}
\label{sec:stress_test}

To isolate the effect of detector degradation, we stress the ranking pipeline by training a reduced-feature (flags‑only) LR. We restrict the model to weakly discriminative TCP flag features, the LR classifier, faced with heavy class imbalance, defaults toward predicting the benign class. As a result, the alert stream and the confidence scores are both degraded. This changes the alert stream itself and produces a low-recall queue (recall 0.1055; F1 0.1786), so the ranking differences become visible (we separately analyze calibration drift with fixed alert sets in Section~\ref{sec:miscalibration}).

Table~\ref{tab:precision_recall_stress} presents the ranking performance under this degraded detector. The results elucidate the disparities in ranking performance.

\begin{table}[h]
\caption{Predicted-alert ranking performance under flags-only LR }
\label{tab:precision_recall_stress}
\begin{tabular}{p{2.8cm} p{2.4cm} p{2.4cm}p{2.4cm}p{2.4cm}}
\toprule
\textbf{Method} & NDCG$_{\text{rel}}$@10$_{\text{pred}}$ & NDCG$_{\text{rel}}$@50$_{\text{pred}}$ & NDCG$_{\text{rel}}$@100$_{\text{pred}}$ & NDCG$_{\text{rel}}$@500$_{\text{pred}}$ \\
\midrule
Severity-Only & 1.0000 & 1.0000 & 1.0000 & 1.0000 \\
Confidence-Only & 0.0040 & 0.0040 & 0.0040 & 0.0041 \\
Weighted Sum & 0.9842 & 0.9009 & 0.8215 & 0.9484 \\
\textbf{Risk-Averse ($\kappa=1$)} & \textbf{1.0000} & \textbf{0.9984} & \textbf{0.9963} & \textbf{0.9911} \\
\bottomrule
\end{tabular}
\end{table}

Our method scores a near-perfect NDCG$_{\text{rel}}$ of 0.9963 under the flags-only stress test. It is particularly striking compared to the baseline methods. To understand this result, consider the ranking task: even with a noisy detector, the true attacks in the predicted-alert queue still have high severity scores (they are real attacks). The Risk-Averse method successfully identifies these high-severity true attacks and ranks them near the top, despite the detector's low confidence in them. When $h_i$ is low, due to its design, the method degrades gracefully rather than catastrophically. The small 0.37\% gap from perfect (0.9963 vs. 1.0000) represents only a few of the ranking errors among the top-100 alerts due to minor misrankings between alerts of similar severity. Other remarks:

\begin{itemize}
\item Severity-Only remains at 1.0 across all $k$'s because NDCG$_{\text{rel}}$@100 assigns zero relevance to benign FPs, so ranking by $c_i$ exactly matches the graded-relevance ordering within the predicted-alert queue.

\item The catastrophic failure of the Confidence-Only method (NDCG$_{\text{rel}}$@100 of 0.0040) illustrates the danger of relying on a single fragile signal. When the detector becomes unreliable, it leads to a ranking that prioritizes high-confidence FPs. Under stress-test conditions, the detector's confidence score no longer matches the actual situation. Consequently, ranking by confidence alone prioritizes a stream of high-confidence FPs, burying the few true attacks and rendering the prioritization useless.

\item Weighted-Sum achieves only 0.8215 at the same cutoff—a 17.85\% degradation because its fixed 0.5 weighting cannot adapt to the now-unreliable confidence signal; it continues to pollute the ranking by giving equal importance to a noisy input. 
\end{itemize}

The key finding of this stress test is that while simple methods may suffice when detectors are perfect, they fail under realistic conditions. Explicitly modeling uncertainty through the fuzzy framework is a practical necessity to maintain robust alert prioritization in the face of real-world detector degradation. The Risk-Averse method provides near-perfect ranking under strong detectors and shows solid robustness under degraded detectors through its principled treatment of multiple uncertainty sources.

\subsection{Confidence-Band Analysis}
\label{subsec:confidence-band}
To isolate where ranking decisions matter most, we partition predicted alerts into three confidence bands (0.3--0.5, 0.5--0.7, 0.7--1.0) and compute NDCG$_{\text{rel}}$@100 within each band. The mid-confidence band (0.5--0.7) is the most diagnostically informative: Confidence-Only drops to 0.3697, while Risk-Averse ($\kappa=1$) remains high at 0.9963. At the broader NDCG$_{\text{rel}}$@500 cutoff, the same mid-band remains stable for Risk-Averse ($\kappa=1$ = 0.9912). Severity-Only and Weighted Sum are near-perfect because confidence varies little within a band. We consider the mid-confidence band as the main way to assess performance, and we include complete band results in the Appendix (Table~\ref{tab:confidence_bands}).

\subsection{Statistical Significance}

To quantify whether the observed differences between methods are statistically meaningful and not due to random chance, we employ a paired bootstrap test on NDCG$_{\text{rel}}$@500 for the CIC-IDS2017 test set with the full-feature LR detector. We use paired resampling because the same test set is evaluated by all methods, making paired comparisons more statistically powerful than the unpaired test, and we use NDCG$_{\text{rel}}$@500 for bootstrap testing because it balances stability and operational focus: @10 and @100 are too sensitive to small rank swaps (high variance). Each comparison gives the mean difference $\Delta$ = NDCG$^A_{rel}$ - NDCG$^B_{rel}$, along with a 95\% confidence interval (CI) and a p-value, the probability of observing a mean NDCG$_{\text{rel}}$@500 difference at least as extreme as the one measured. Positive $\Delta$ indicates that Method A ranks higher-severity attacks better; negative $\Delta$ indicates Method B performs better.
\begin{table}[h]
\caption{Paired bootstrap results on NDCG$_{\text{rel}}$@500}
\label{tab:bootstrap_tests}
\begin{tabular}{l l c c c}
\toprule
\textbf{Method A} & \textbf{Method B} & \textbf{$\Delta$} & \textbf{95\% CI} & $p$ \\
\midrule
Confidence-Only & Risk-Averse ($\kappa=1$) & -0.2265 & [-0.2555, -0.1909] & 0.0001 \\
Severity-Only & Risk-Averse ($\kappa=1$) & 0.0016 & [-0.0014, 0.0029] & 0.1620 \\
Weighted Sum & Risk-Averse ($\kappa=1$) & 0.0022 & [0.0008, 0.0031] & 0.0260 \\
\bottomrule
\end{tabular}
\end{table}

Table~\ref{tab:bootstrap_tests} shows that Risk-Averse significantly outperforms Confidence-Only; shows no statistically reliable difference versus severity baseline, and almost as good as Weighted Sum with a tiny effect size. These conclusions are stable across @1{,}000, @5{,}000, and @10{,}000 bootstrap resamples.

\subsection{Auxiliary Detector Families}
To assess sensitivity to the detector family, we ran auxiliary experiments with a supervised Random Forest and an unsupervised Isolation Forest. The Random Forest backend yields near-perfect rankings for severity-based methods because the detector is strong and confidence scores saturate; the comparison therefore collapses, as expected. The Isolation Forest backend highlights a different failure mode: its anomaly scores are not calibrated as attack probabilities, so Confidence-Only collapses even after normalization, while Risk-Averse remains stable because it continues to anchor rankings on severity. Full results are shown in Appendix~\ref{subsec:aux-detectors}.

\subsection{Robustness Under Detector Miscalibration}
\label{sec:miscalibration}
Real-world detectors often become miscalibrated due to domain shift (when network traffic patterns change), seasonal variations (when attack patterns change over time), and training data drift (when the historical data used to train the detector no longer reflect current conditions). Unlike the stress test, which changes the detector and thus the alert stream, in this section we isolate the calibration drift by keeping the alert set fixed and perturbing only the confidence scores $p_i$. This let us measure how sensitive each ranking method is to miscalibration without conflating the effect with changes in detection coverage. We test the following three scenarios:

\paragraph{Scenario 1: Optimistic Bias (Detector Overconfident)} -- $p_i \times 1.15$: simulates a detector that systematically overestimates its confidence. This happens when a detector is trained on past data, but encounters a new attack distribution with slightly different characteristics. We simulate this by scaling confidence scores: $p'_i = min(1.0, p_i \times 1.15)$, capping at 1.0 to maintain validity. The results show minimal degradation across all methods:

\begin{itemize}
    \item Severity-Only: 0.00\%
    \item Confidence-Only: $-$0.13\%
    \item Weighted Sum: $-$1.53\%
    \item Risk-Averse: 0.00\%
\end{itemize}

The key insight is that Risk-Averse's height-capping mechanism ($h_i \leq 1.0$) naturally bounds the effect of overconfidence. When $p_i$ is scaled up, the height $h_i$ saturates at 1.0, limiting the damage to the ranking. Conversely, the Weighted Sum's constant 0.5 weighting permits the inflated confidence to distort the ranking, resulting in a 1.53\% decline. Consequently, this illustrates that Risk-Averse's fuzzy framework inherently offers resilience against systematic overconfidence.

\paragraph{Scenario 2: Pessimistic Bias (Detector Underconfident)} -- $p_i \times 0.85$: simulates a uniformly conservative detector by scaling all confidence scores down: $p'_i = p_i \times 0.85$. Uniform scaling preserves ordering (0.00\% change for all methods). This finding validates that our normalization strategy, which scales confidence to match CVSS range, successfully absorbs uniform calibration bias.

\paragraph{Scenario 3: Random Drift (+0.2 Noise)} – $p_i + N(0, 0.2)$: simulates stochastic noise in confidence scores, representing random fluctuations due to detector instability, network variability, or measurement error. We add a Gaussian noise: $p'_i = clip(p_i + N(0, 0.2), 0, 1)$. The results show a clear differentiation between methods: noise harms Confidence-Based methods and Risk-Averse remains stable because the logarithmic term reduces the effect of noise, and the uncertainty term $\sigma_i$ provides a damping effect.

\begin{itemize}
    \item Severity-Only: 0.00\%
    \item Confidence-Only: $-$2.93\%
    \item Weighted Sum: $-$0.71\%
    \item Risk-Averse: +1.92\%
\end{itemize}

Across all scenarios, Risk-Averse is at least as robust as the baselines. Height-capping bounds overconfidence, normalization absorbs uniform bias, and log-scaling with uncertainty damps stochastic noise.

\section{Conclusion}
\label{sec:conclusion}

Alert fatigue represents a significant impediment within security operations, as analysts are inundated with numerous daily alerts, the majority of which are FPs. This research tackles this issue by introducing a framework designed for alert prioritization. The framework explicitly incorporates three sources of uncertainty: threat severity, detection confidence, and organizational risk attitude.

Our key contributions include: (1) introducing subnormal Gaussian fuzzy numbers to IDS alert prioritization, enabling principled uncertainty modeling, (2) providing a novel framework that allows organizations to tune their security posture through the risk-attitude parameter $\kappa$, (3) validating that our framework is near-parity with baselines on strong detectors but substantially robust when detectors degrade with different attack distributions on CIC-IDS2017 and NSL-KDD datasets, (4) showing that Risk-Averse maintains 0.9963 NDCG$_{\text{rel}}$@100 while alternatives degrade to 0.0040–0.8215, achieving greater robustness, and (5) revealing that the mid-confidence band (0.5–0.7) is where explicit uncertainty modeling provides the greatest value.


In practice, such as in a typical SOC with 10,000 alerts/day and 80\% FPs, even a 10\% improvement in ranking quality frees analysts to focus on critical threats. Our method achieves this while maintaining robustness to realistic detector degradation.

Statistical significance is validated through paired bootstrap testing (p < 0.05), and robustness is confirmed across detector families (Logistic Regression, Random Forest, Isolation Forest) and realistic miscalibration scenarios (overconfidence, underconfidence, stochastic noise).

Regarding limitations and future research: we assume CVSS scores are available for all attacks and novel attacks may lack scores, thus evaluation on public datasets should be supplemented with real-world SIEM data. In addition, parameter $\kappa$ is fixed; therefore, future work should develop adaptive learning methods for this parameter. Future work will also incorporate temporal dynamics to capture evolving attack patterns and detector performance drift, enabling adaptive ranking under non‑stationary conditions.


\appendix

\section{Supplementary Tables}

\subsection{Experimental Protocol}
This section provides a detailed, step-by-step recipe that shows how we conducted our experiments for the sake of transparency, validation, and reproducibility. Having prepared our dataset, the phases below are followed.

\subsubsection{Training Phase}
In this phase, we:
\begin{enumerate}
    \item train the baseline IDS (LR) on the training set (Mon/Tue data) with solver \texttt{saga}, $C=1.0$, max iterations $=1000$, and class weights set to \texttt{balanced},
    \item if the day-based split is attack-sparse (<5\%), fall back to a stratified 50/20/30 split
    \item calibrate predicted probabilities on the validation set using a sigmoid calibrator and evaluate the baseline IDS on the validation set (Wed data),
    \item for each attack class $j$ compute detection confidence metrics (Precision, Recall, F1) for each attack class, and
    \item derive height values $h_j$ from the smoothed F1 and store them in a lookup table to use during alert ranking.
\end{enumerate}

\subsubsection{Testing Phase} In this phase, for each test set (Thu/Fri data), we:

\begin{enumerate}
    \item run the baseline IDS on the test set to generate alerts,
    \item for each alert, extract the attack type and raw IDS confidence,
    \item for each ranking method (Severity-Only, Confidence-Only, Weighted Sum, Risk-Averse):
    \begin{enumerate}
        \item compute the ranking score for each alert;
        \item sort alerts by ranking score;
        \item compute evaluation metrics
    \end{enumerate}
    \item compare the ranking methods across all metrics
\end{enumerate}

\subsubsection{Parameter Tuning}

We sweep $\kappa \in \{0, 0.5, 1, 1.5, 2\}$ and present the sensitivity analysis in Appendix~\ref{subsec:parameter_tune}; main results use $\kappa=1$ as a standard setting rather than tuning per dataset.

\subsubsection{Statistical Significance Testing}

To quantify whether differences between methods are statistically meaningful, we perform a paired bootstrap test on NDCG$_{\text{rel}}$@500 computed from the top-500 ranked alerts:

\begin{enumerate}
    \item For each test set, take the top-500 alerts from each method (based on rank) and form graded relevance vectors
    \item Compute NDCG$_{\text{rel}}$@500 for each method and bootstrap the mean difference using paired resampling over rank positions
    \item Report bootstrap p-values and 95\% confidence intervals for the mean difference in NDCG$_{\text{rel}}$@500
    \item Consider the differences significant at $p < 0.05$
\end{enumerate}

\subsection{Height Parameter and Risk-Attitude Sensitivity}
\label{subsec:parameter_tune}
In the main text, we present results with the default setting $\kappa=1$; here we provide sensitivity examples to show how rankings shift under higher risk aversion.
To demonstrate the value of the height parameter (detection confidence), we used three real DoS alerts from the CIC-IDS2017 test set with low, medium, and high heights by default $\kappa=1$. The low-height alert (alert\_id 353856, $h_i=0.3796$) is ranked lower than the medium and high-height alerts by the method, while the confidence-only baseline places it much lower overall. Table \ref{tab:height_impact} also lists the corresponding severity core $c_i$ and spread $\sigma_i$ values.
\begin{table}[h]
\caption{Impact of Height Parameter on Alert Ranking (CIC-IDS2017 test set)}
\label{tab:height_impact}
\begin{tabular}{p{1.3cm} p{1.1cm} p{1.1cm}p{1.1cm}p{1.1cm}p{1.1cm}p{2.0cm}p{1.9cm}}
\toprule
\textbf{Alert ID} & \textbf{Type} & $h_i$ & $p_i$ & $c_i$ & $\sigma_i$ & \textbf{Risk-Averse Rank} & \textbf{Confidence-Only Rank} \\
\midrule
353856 & DoS & 0.3796 & 0.3796 & 7.2504 & 1.4706 & 24463 & 179313 \\
192641 & DoS & 0.7992 & 0.9872 & 7.3872 & 1.4267 & 5195 & 60304 \\
230833 & DoS & 0.7992 & 1.0000 & 7.8000 & 1.2480 & 403 & 1345 \\
\bottomrule
\end{tabular}
\end{table}

When an alert has low detection confidence, our method deprioritizes it (rank 24463); a confidence-only baseline, driven solely by $p_i$, places it even lower (rank 179313). This demonstrates the value of explicitly modeling detection confidence through the height parameter.

We also show $\kappa$ sensitivity for three DoS alerts with low, medium, and high heights to show how the risk-attitude parameter shifts rankings. As $\kappa$ increases, the low-height alert is penalized more, while medium- and high-height alerts shift only modestly (Table \ref{tab:kappa_impact}). This flexibility allows organizations to tune their security posture: aggressive organizations can use $\kappa=0$ to focus on high-severity threats, while conservative organizations can use $\kappa \geq 1$ to focus on high-confidence threats.
\begin{table}[h]
\caption{Impact of Risk-Attitude Parameter $\kappa$ on Alert Ranking. As $\kappa$ increases, the ranking shifts. Specifically, the alert with the highest severity but lowest confidence (Rank 9467 when $\kappa$=0) is moved down the list. At the same time, the alert with lower severity but higher confidence (Rank 403 when $\kappa$=1) is moved up. This change illustrates the trade-off between severity and confidence.}
\label{tab:kappa_impact}
\begin{tabular}{llccccccc}
\toprule
\textbf{Alert ID} & \textbf{Type} &  $h_i$ & $c_i$ & $\sigma_i$ & $\kappa=0$ & $\kappa=1$ & $\kappa=2$ \\
\midrule
353856 & DoS low $h_i$ &  0.3796 & 7.2504 & 1.4706 & Rank: 9467 & Rank: 24463 & Rank: 41807 \\
192641 & DoS med $h_i$ & 0.7992 & 7.3872 & 1.4267 & Rank: 7436 & Rank: 5195 & Rank: 5426  \\
230833 & DoS high $h_i$ & 0.7992 & 7.8000 & 1.2480 & Rank: 946 & Rank: 403 & Rank: 675 \\
\bottomrule
\end{tabular}
\end{table}

\subsection{Contextual Criticality (Heartbleed)}
To show that more critical instances of the same attack type are prioritized more, we use two Heartbleed alerts from the test set with different contextual factors under the default setting $\kappa=1$. The higher-criticality alert (larger $\text{cf}_i$) is ranked far above the lower-criticality alert, indicating that contextual criticality remains decisive even when risk adjustment is applied (Table \ref{tab:criticality_heartbleed}).
\begin{table}[h]
\caption{Contextual criticality effect for Heartbleed alerts (CIC-IDS2017 test set, $\kappa=1$)}
\label{tab:criticality_heartbleed}
\begin{tabular}{llccccc}
\toprule
\textbf{Alert ID} & \textbf{Type} & $\text{cf}_i$ & $c_i$ & $h_i$ & $p_i$ & \textbf{Risk-Averse Rank} \\
\midrule
295490 & Heartbleed & 0.2242 & 2.1792 & 0.0002 & 0.0002 & 849172 \\
57807 & Heartbleed & 0.9838 & 9.8384 & 0.5000 & 0.9575 & 13886 \\
\bottomrule
\end{tabular}
\end{table}

\subsection{Confidence-Band Analysis}
Table~\ref{tab:confidence_bands} provides the full three-band breakdown for the flags-only stress test with $\kappa=1$. The mid-confidence band (0.5--0.7) is the most discriminative region, while the low- and high-confidence bands show compressed differences. This table complements the main-text focus on mid-confidence alerts as the primary diagnostic view.
\begin{table}[h]
\caption{Confidence-band NDCG$_{\text{rel}}$@100 under flags-only LR}
\label{tab:confidence_bands}
\begin{tabular}{p{2.8cm} p{2.2cm} p{2.2cm}p{2.2cm}}
\toprule
\textbf{Method} & NDCG$_{\text{rel}}$@100 (0.3--0.5) & NDCG$_{\text{rel}}$@100 (0.5--0.7) & NDCG$_{\text{rel}}$@100 (0.7--1.0) \\
\midrule
Severity-Only & 0.9802 & 1.0000 & 1.0000 \\
Confidence-Only & 0.2692 & 0.3697 & 0.0053 \\
Weighted Sum & 0.9802 & 1.0000 & 1.0000 \\
\textbf{Risk-Averse ($\kappa=1$)} & 0.9982 & 0.9963 & 0.9601 \\
\bottomrule
\end{tabular}
\end{table}

\subsection{Auxiliary Detector Families}
\label{subsec:aux-detectors}
We present auxiliary results for a supervised Random Forest and an unsupervised Isolation Forest to probe detector-family sensitivity. Each cell includes NDCG$_{\text{rel}}$@10$_{\text{pred}}$ / NDCG$_{\text{rel}}$@100$_{\text{pred}}$ for the predicted-alert queue. Random Forest is a strong supervised backend, so severity-based rankings are near-perfect. Isolation Forest uses anomaly scores that are not calibrated probabilities, so Confidence-Only performs poorly, while Risk-Averse remains stable by anchoring on severity.
\begin{table}[h]
\caption{Auxiliary detector results (NDCG$_{\text{rel}}$@10$_{\text{pred}}$ / NDCG$_{\text{rel}}$@100$_{\text{pred}}$)}
\label{tab:aux_detectors}
\begin{tabular}{p{1.2cm}p{1.8cm} p{2.0cm}p{2.0cm}p{2.0cm}p{2.0cm}}
\toprule
\textbf{Detector} & \textbf{Setting} & \textbf{Severity-Only} & \textbf{Confidence-Only} & \textbf{Weighted Sum} & \textbf{Risk-Averse ($\kappa=1$)} \\
\midrule
RF & Full features & 1.0000 / 1.0000 & 0.5794 / 0.5658 & 1.0000 / 1.0000 & 1.0000 / 1.0000 \\
RF & Flags-only & 1.0000 / 1.0000 & 0.3697 / 0.3697 & 1.0000 / 1.0000 & 1.0000 / 0.9963 \\
IF & Full features & 0.8689 / 0.9715 & 0.0000 / 0.0000 & 0.9029 / 0.9732 & 1.0000 / 1.0000 \\
IF & Flags-only & 0.8689 / 0.9715 & 0.0223 / 0.0223 & 0.9676 / 0.9855 & 1.0000 / 1.0000 \\
\bottomrule
\end{tabular}
\end{table}

\subsection{Parameter Sensitivity}
\label{app:sensitivity}
We varied height-smoothing parameters and globally scaled uncertainty factors to evaluate sensitivity of Risk-Averse ranking (full-feature LR, CIC-IDS2017). Table~\ref{tab:parameter_sensitivity} shows that NDCG$_{\text{rel}}$@10$_{\text{pred}}$ stays within 0.9992--0.9995 and NDCG$_{\text{rel}}$@100$_{\text{pred}}$ within 0.9862--0.9916 across all settings. The largest shifts occur under global scaling of uncertainty factors; $\alpha$, $h_{\min}$, and $h_{\max}$ have negligible effect in this range.
\begin{table}[h]
\caption{Risk-Averse sensitivity to height smoothing and uncertainty scaling (NDCG$_{\text{rel}}$@10$_{\text{pred}}$ / NDCG$_{\text{rel}}$@100$_{\text{pred}}$).}
\label{tab:parameter_sensitivity}
\begin{tabular}{lccc}
\toprule
\textbf{Parameter} & \textbf{Value} & \textbf{NDCG$_{\text{rel}}$@10$_{\text{pred}}$} & \textbf{NDCG$_{\text{rel}}$@100$_{\text{pred}}$} \\
\midrule
$\alpha$ & 0.50 & 0.9994 & 0.9914 \\
$\alpha$ & 0.70 & 0.9994 & 0.9890 \\
$\alpha$ & 0.90 & 0.9994 & 0.9890 \\
$\alpha$ & 0.95 & 0.9994 & 0.9890 \\
$h_{\min}$ & 0.01 & 0.9994 & 0.9890 \\
$h_{\min}$ & 0.05 & 0.9994 & 0.9890 \\
$h_{\min}$ & 0.10 & 0.9994 & 0.9890 \\
$h_{\max}$ & 0.90 & 0.9994 & 0.9890 \\
$h_{\max}$ & 0.95 & 0.9994 & 0.9890 \\
$h_{\max}$ & 0.99 & 0.9994 & 0.9890 \\
$\text{uf}_i$ scale & 0.80$\times$ & 0.9995 & 0.9916 \\
$\text{uf}_i$ scale & 1.00$\times$ & 0.9994 & 0.9890 \\
$\text{uf}_i$ scale & 1.20$\times$ & 0.9992 & 0.9862 \\
\bottomrule
\end{tabular}
\end{table}

\subsection{Multi-Dataset Validation: NSL-KDD}
We also evaluated the framework on NSL-KDD \cite{Tavallaee2009} as a generalization check across datasets with different attack distributions. Using the full-feature LR detector, ranking methods are near-identical, consistent with the CIC-IDS2017 pattern where strong detectors compress differences. This supports the view that meaningful separation appears primarily when the alert stream is noisy or uncertain.

\section{Implementation Details}

\subsection{Software and Libraries}

\begin{itemize}
    \item {\textbf{Language}}: Python 3.8+
    \item {\textbf{Machine Learning}}: scikit-learn 1.0.2 (Logistic Regression, metrics)
    \item {\textbf{Data Processing}}: pandas 1.3.5, numpy 1.21.6
    \item {\textbf{Visualization}}: matplotlib 3.5.1, seaborn 0.11.2
    \item {\textbf{Statistical Testing}}: numpy/scipy (paired bootstrap)
\end{itemize}

\subsection{Experiment Configuration}
\begin{itemize}
    \item {\textbf{Baseline IDS}}: Logistic Regression (solver \texttt{saga}, $C=1.0$, max iterations $=1000$, class weights \texttt{balanced}) with sigmoid probability calibration.
    \item {\textbf{Risk Parameters}}: $\kappa \in \{0, 0.5, 1, 1.5, 2\}$; height smoothing $\alpha=0.9$, $h_{\min}=0.05$, $h_{\max}=0.95$; instance height $h_i=\min(h_{\text{class}}, p_i)$.
    \item {\textbf{Stress Test}}: flags-only feature subset (columns containing ``flag''), minimum 5 features.
    \item {\textbf{Confidence Bands}}: $[0.3, 0.5)$, $[0.5, 0.7)$, $[0.7, 1.0]$.
    \item {\textbf{Random Seed}}: seed=42 for numpy/python RNG and data splitting.
    \item {\textbf{Auxiliary Detectors}}: Random Forest (supervised) and Isolation Forest (unsupervised) runs for detector-family sensitivity, shown in Appendix~\ref{subsec:aux-detectors}.
\end{itemize}

\subsection{Reproducibility}

To ensure reproducibility:

\begin{itemize}
    \item Random seeds are fixed (seed=42) for data splits and RNG-dependent steps
    \item Dataset preprocessing and split logic (including CIC-IDS2017 fallback when the train split has $<5\%$ attacks) is documented
    \item Hyperparameters for the baseline IDS are specified
    \item Attack-type mapping is centralized with an optional override CSV
    \item Results, plots, and dataset profiles are saved in \textbf{results/} folder
\end{itemize}

\bibliographystyle{unsrtnat} 
\bibliography{references}

\end{document}